\documentstyle[psfig,frontieres,art10]{article}
\begin{document}
\def\ee{\end{equation}}
\def\be{\begin{equation}}
\heading{Strategy to detect the gravitational radiation counterpart of 
$\gamma$-ray bursts\footnote{to appear in the Proceedings of the
Second Workshop on Gravitational Wave Data Analysis (Orsay,
13-15 November 1997).}}

\author{
Silvano Bonazzola${}^1$, Eric Gourgoulhon${}^1$} 
 
\address{
D.A.R.C. (UPR 176 du CNRS), 
Observatoire Paris-Meudon, 92195 Meudon Cedex, France}

\begin{abstract} 
Both observational and theoretical rates of binary neutron star coalescence 
give low prospects for detection of a single event by the
initial LIGO/VIRGO interferometers. However, by utilizing at the best
all the a priori information on the expected signal, a positive detection can
be achieved.  This relies on the
hypothesis that $\gamma$-ray bursts are the electromagnetic signature of
neutron star coalescences. The information about the direction of the source 
can then be used to add in phase the signals from different detectors in order 
(i) to increase the signal-to-noise ratio and (ii) to make the noise more
Gaussian. Besides, the 
information about the time of arrival can be used to drastically decrease the
observation time and thereby the false alarm rate. 
Moreover the fluence of the $\gamma$-ray emission gives some information 
about the amplitude of the gravitational signal. One can then add the signals
from $\sim 10^4$ observation boxes ($\sim$
number of $\gamma$-ray bursts during 10 years) to yield a positive 
detection. 
Such a detection, based on the Maximum a Posteriori 
Probability Criterium, is a minimal one, in the sense that no 
information on the position and time of the events, nor on any parameter of 
the model, is collected. 
The advantage is that this detection requires an improvement of the
detector sensitivity by a factor of only
$\sim 1.5$ with respect to the initial LIGO/VIRGO interferometers, and that,
if positive, it will confirm the $\gamma$-ray burst model.  
\end{abstract} 
\vskip 0.8cm

A widely spread out gamma-ray burst model is related with the coalescence
of two  neutron stars (N.S.) at cosmological distance  \cite{mosco}. 
In this model, the gravitational energy liberated
during the disruption of the less massive star of a binary system of N.S.
is transformed into electromagnetic energy and radiated in the X,$\gamma$
range \cite{bona}. 
\medskip

No direct evidence of the validity of this model exits until now.
Only energy budget considerations and an estimated rate of coalescences of N.S.
computed in the frame of stellar evolution theory, confort the idea that
at least a large fraction of $\gamma$-ray bursts are generated by the above 
mechanism \cite{Yungel}.
\medskip

 In this communication, we want to show that  
 the class of a typical noise of $10^{-23}/\sqrt{\rm  Hz} $ G.W. detectors is 
sensitive
 enough to detect, after a few years of  observation, the associated G.W. 
emission of
the coalescence. The strategy proposed here consists in using at the best
all the a priori information and hypothesis of the model and to ``summ''
the signals of many events.
\medskip

Historically, coalescing N.S. were considered as the most promising
source of detectable G.W. A detection rate of few events/year was estimated
in an (to much) optimistic case (for a review on the subject see \cite{bona}).
Since then, there is widely spread out consensus that the coalescing
rate of N.S. is one per year in a sphere of $200\  {\rm Mpc}$ of radius.
$\gamma$-ray burst have a repetition rate $10$ times lower, and a focusing
mechanism must be invocated in order to explain this discrepancy
\cite{Yungel}.
\medskip

 The predicted sensitivity of the first VIRGO detector will allow us
to detect a coalescence of N.S. at a distance of $27 {\ \rm Mpc}$ with a 
signal/noise (S/N)
ratio of $7$ \cite{Hello}
 (the figure $7$ was choosen in view to have one false alarm per
year under the hypothesis that the noise is Gaussian). Therefore the  
detection rate of coalescing N.S. will result to be $0.5$ event per 
century (!) or $10$ times lower if no beaming mechanism exist in the 
$\gamma$-ray burst emission.
\medskip

\begin{figure} 
\centerline{\vbox{ 
\psfig{figure=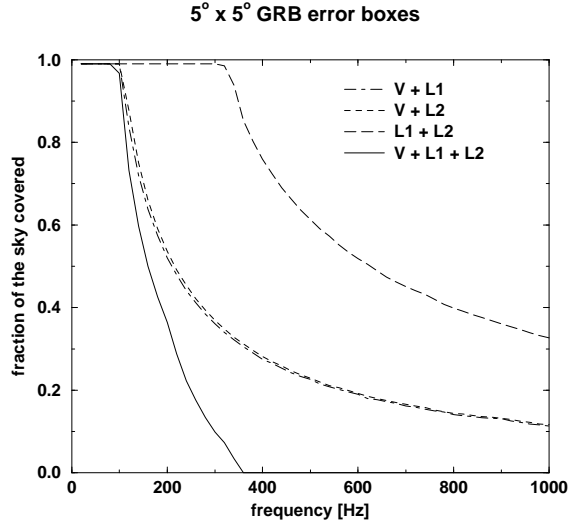,height=7.0cm} 
}} 
\caption[]{Fraction of the sky covered when phasing the signal of 
different detectors (L1 = LIGO 1, L2 = LIGO 2, V = VIRGO), in the case
where the direction of the source is known with an accuracy of $5^o$.
A portion of the sky is said to be covered if the phase lag 
between different detectors resulting
from the incurate knowledge of the source position is lower than $\pi/2$.}
\end{figure} 

\begin{figure} 
\centerline{\vbox{ 
\psfig{figure=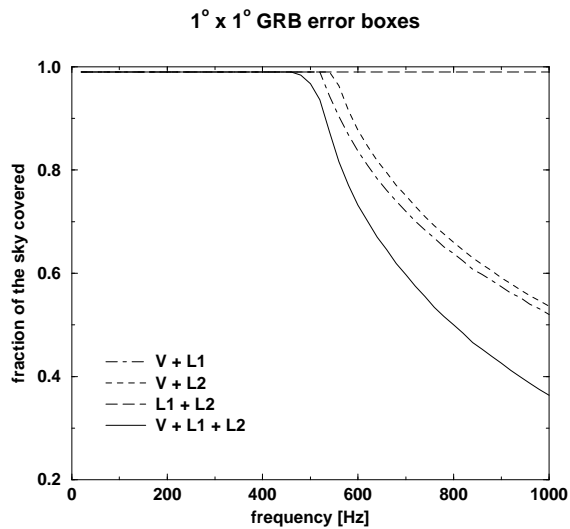,height=7.0cm} 
}} 
\caption[]{Same as Fig.~1 when the direction of the source is known with an 
accuracy of $1^o$ instead of $5^o$.}
\end{figure}

 In view to increase the detection rate it is convenient to add in phase the
signal of different detectors. In fact if $N$ is the number of detectors, the 
S/N  increases as
the $\sqrt{N}$ and the detection rate as $N^{3/2}$. Moreover, adding different
signals has the big advantage to make the signal more Gaussian (Central limit
theorem !) and therefore to decrease the false alarm rate.
\medskip

 Adding in phase the outputs of different detectors is easy if the position
of the source in the sky is known with a sufficient accuracy. This is the case
of an optical or radio detection of the source (pulsars or supernovae).

 For $\gamma$-ray bursts, the error box of the recent BATSE experiment
 aboard the Compton-GRO satellite is about $5^{0}\times 5^{0}$. This accuracy
in 
the determination of the position of the source in the sky will allow the
phasing of the signals from $3$ detectors (VIRGO, LIGO 1 and LIGO 2) 
for signals whose frequencies are lower than $200$ {\rm Hz} (See Fig 1). 
Figure~2 shows that
reducing the error boxes by a factor of $5$ will be enough to phase
the output of $3$ detectors. It is quite possible that the new generation of
$X$ and $\gamma$-ray detectors will improve the localization accuracy. In
our opinion, the Gravitational Wave Community can play an important role
in convincing the $\gamma$-ray Community of the importance of improving the
accuracy of positioning the sources.
\medskip

 The hypothesis that $\gamma$-ray bursts are generated by coalescing N.S.
allows  us to suppose that the frequency of the gravitational radiation  
sweeps the frequency window of the detectors a few seconds before the begining
of the $\gamma$-ray emission; moreover, we suppose that the strength of the
gravitational signal and  the fluence of the electromagnetic one are correlated
with the distance of the source. These quite natural hypotheses will be considered as a priori informations and will be used in computing the a posteriori
probability density of detection of the signal.
Other a priori informations will be enumerated (and used) in what follows.
\medskip

 The Maximum a Posteriori Probability Criterium (M.P.P.C.) \cite{kend} 
will be the tool used
 in view  to utilize at the best all the a prior information given by the 
model. Let us  consider a simple example to show how the M.P.P.C. works.    
\medskip

  Consider a sinusoidal signal of length $T$: $y(t)=a\times\cos (\omega t+\phi)$ 
in presence of a white Gaussian noise $n(t)$.
The sampled signal is:
\begin{equation}
y_i=a\times\cos (\omega t_i+\phi)+n_i,\quad i=1,2...,N \ ,
\end{equation}
where $N$ is the number of sampling points: $N=T\times\nu_s$, $\nu_s$ being the
sampling frequency.
  The density probability of having $N$ $y_i$ values is
\begin{equation}
P(y_1,y_2,...y_N | a,\phi)=\frac{1}{(\sigma \sqrt{2\pi)^{N}}} 
\exp{(-\sum_{i=1}^{N}
(y_{i}-a\cos (\omega t_{i}+\phi))^{2}/2\sigma^{2})} \ .
\end{equation}
Let us suppose that the phase $\phi$ is known. In this case the maximum
likelihood criterium allows us to estimate the amplitude of the signal $a$.
\medskip

The best estimation of $a$ is given by solving the equation $ $
 $ \partial_{a}{P(y_1,y_2,...y_N | a,\phi)} =0$,
from which we obtain the well
 known result $a=2/N \sum_{i=1}^{N} y_{i}\cos (\omega t_{i}+\phi )$
and the Signal/Noise ratio reads 
\begin{equation}
S/N=aT/B \ , 
\end{equation}
where $T$ is the time length of the signal (\rm sec.) and $B$ is the 
instrumental noise per $\sqrt{\rm Hz}$.
\medskip

Consider now the little  more complicated case, in which the phase $\phi$ is not 
 known. We have two possible strategies: the first one consists in estimating
simultaneously $a$ and $\phi$ by solving the system 
$ \partial_{a}{P(y_1,y_2,...y_N | a,\phi)} =0$, 
$ \partial_{\phi }{P(y_1,y_2,...y_N | a,\phi)} =0 $. 
 In this way we obtain one  value for $a$ and a value for $\phi$. 
This gives too much 
information because we do not need to know $\phi$.
 The alternative strategy consists in maximizing the {\it posteriori probability
density} defined by $\int_{0}^{2\pi }  P(\phi )P(y_1,y_2,...y_N |a,\phi)d\phi$
where $P(\phi) $ is the a priori probability density of the phase. Because we
have no  information on the phase 
(except that it spans the interval $[0,2\pi[$)
 we apply the so-called {\it Equipartition of the Ignorance Principle} 
\cite{kend},\cite{levi},\cite{kend2} to set 
$P(\phi )=const.=1/2\pi $.
 After integration, the amplitude $a$ is found by the maximum of the function 
$\exp{a}\times I_{0}(a)$ where $I_{0}$ is the Bessel function of $0$
order \cite{levi1}. 
The asymptotic value of $a$  for $ S/N \rightarrow\infty$ is
 $a=\sqrt{c^{2}+s^{2}}$ where 
$c = 1/N\sum_{i=1}^{N} y_{i}\cos(\omega t_{i}) $ and 
$s = 1/N\sum_{i=1}^{N} y_{i}\sin (\omega t_{i})$,
i.e. the value of $a$ is given by the power spectrum of the signal.
The $S/N$ is
\begin{equation}
S/N=aT/(\sqrt{2}\times{B}) \ ,
\end{equation}
 i.e. $1/\sqrt{2}$ times worse
than in the previous case in which the phase was known. 
The factor $\sqrt{2}$
is the price to pay for our ignorance. 
If more then one ``box'' exist, say $J$ boxes, then
the M.P.P.C. tell us that we have to maximize the a posteriori probability given
by the product of the density probability of each box:
\begin{equation}
P_1(y_1,y_2,...y_N | a,\phi)P_2(y_2,y_2,...y_N | a,\phi)...
P_J(y_1,y_2,...y_N | a,\phi)
\end{equation}
It is easy to show that, roughly speaking, the signal-to-noise ratio is 
increased by a factor $J^{1/4}$.
\medskip

In the real case, i.e. the detection of the signal from coalescing N.S., the
probability density distribution $P(y_1,y_2,...y_N |a, \phi,m_1,m_2,i..)$
depends on more variables than $a$ and $\phi$,
namely the masses $m_1$ and $m_2$ of
the two N.S., the angle $i$ between the orbital plane and the line of sight,
etc... 
The a priori informations that we have about these parameters must be used
(for example, we know that the masses of the two N.S. cannot be arbitrary, but
lie between, say, $M_1=0.2 M_\odot$ and
 $ M_2= 3 M_\odot $, and so on). Therefore, the a 
posteriori probability density reads
\begin{eqnarray}
\int_{0}^{2\pi }\int_{M_1}^{M_2}\int_{M_1}^{M_2}\int_{0}^{\pi/2 }...\int &&  
 P(m_1)P(m_2)P(i)P(\phi )\ldots \times \nonumber \\
 && \times P(y_1,y_2,...y_N|a,\phi,m_{1},m_{2},i..) 
    \, d\phi \, dm_1 \, dm_2 \, di...
\end{eqnarray}
\medskip

Under the conservative hypothesis that gravitational radiation observations
must start, say, one minute before the electromagnetic counterpart of the 
$\gamma$-ray burst (the time at which the two N.S. merge is not 
precisely known either for observational reasons and for the lack of a 
reliable model), the total
observation time is decreased by a factor of $\sim 2\times10^{-3}$
(1 minute/(number of minutes per day)$\times$(number of $\gamma$-ray bursts 
per day)).
Therefore the false alarm probability is reduced of the same factor
(vice-versa
$S/N$ is reduced to 6 for the same false alarm probability). 
Note that the idea of using the information given by the timing and 
direction of the $\gamma$-ray bursts to increase the detection rate
has been already developed by
Kochanek \& Piran \cite{Kocha}. 

With five 
detectors ($2+2$ for LIGO 1 and LIGO 2 and $1$ for VIRGO) the distance at 
which a N.S. coalescence can be detected with a $S/N$ of $7$ is increased 
by factor $\sqrt{5}$,
 i.e. $60$ ${\rm Mpc}$ or $420$ ${\rm Mpc}$ for $S/N=1$.
If the coalescence happens at a distance of, say, $4.2$ ${\rm Gpc}$, the
$S/N$ ratio will be $0.1$. Taking into account that in 10 years of 
observation time, $10^{4}$ $\gamma$-ray bursts will be observed (The 
$\gamma$-ray bursts rate is about 3 per day), we shall dispose of $10^{4}$ 
``boxes''
with a gravitational  $S/N$ ratio of ${\sim 0.1}$ in each box. 

By ``adding'' the 
signal of the $10^{4}$ boxes, as explained in the above example, the 
 $S/N$ ratio will be increased by factor of 10 
($=(10^{4})^{1/4}$), i.e. $S/N=1$. 
Note that this is a pessimistic estimation: in fact
 not all
$\gamma$-ray bursts are situated at $4.2$ $\rm{Gpc}$. Under the hypothesis of
an uniform spatial distribution, the average $\gamma$-burst distance
 is reduced by a factor $\sqrt{3}$ and therefore the signal-to-noise
ratio increases by a factor of$\sqrt{3}$: $S/N=1.73$. This is not the end 
of the story: in fact we have not used yet the hypothesis that the intensity
 of the
gravitational signal is correlated with the electromagnetic one  (more precisely
with its fluence).

 The  operation of ``adding'' the different boxes
must be done by weighing the boxes with a weight proportional to the 
fluence of
the $\gamma$-ray bursts. The final result is $S/N=5.5$. This is 
not enough to have an acceptable false alarm probability; in fact a $S/N\ge
7$ is required to have a false alarm probability $\le 10^{-3}$.

It appears that
an improvement by a factor $\ge 1.3$ of the sensibility of the detectors
will be enough to test the hypothesis that $\gamma$-ray bursts are generated
by coalescing  N.S.

The above results must be considered as preliminary ones. In fact we have
supposed that the detectors are aligned (optimistic hypothesis); in real
calculation the directionality of the detectors should be taken into account
and used to weigh the signal of each box. The $S/N$ ratio should be improved
a little bit.
\bigskip

\noindent{\bf CONCLUSIONS} 
\medskip

There is a general agreement that N.S. coalescence rate  is $\sim 1$
per $\rm year$ within a sphere of $200$ $\rm Mpc$. Taking into account
the incertitude of the above estimation (a factor of 2 - 3, 
L.~Yungelson private communication), this value is
in a good agreement with the observed rate of $\gamma$-ray bursts 
in the same volume ($0.1$
per ${\rm year}$). A possible beaming of a factor $10$ is plausible
and is invocated in view to eliminate the discrepancy between the theoretical
model and the observational data.

A more important beaming of the $\gamma$-ray emission during the merging of 
two N.S. would increase the coalescence
rate. It is however  quite difficult to imagine which physical mechanism would 
decrease
the isotropy of the electromagnetic radiation: the velocity
of the two merging N.S. is indeed about half the velocity of light
and therefore the aberration effects are not strong.

Recently, Wilson et al. \cite{wil}
have found by numerical simulations that the more massive of the  two N.S.
can collapse and form a black hole before the merging if its mass is
close to the critical one. Even in this unlikely particular case the 
electromagnetic radiation is not suppressed: 
in fact the less massive star will be destroyed
by the tidal forces of the just born black hole. 
\medskip

An improvement of the detectors sensitivity by a factor of at list $10$
is required in order to detect a few individual events per year with a 
reasonable false alarm probability. 
Nobody knows when and how the thermal noise
of the mirrors will be reduced by  such an  important factor
(one of us (S.B.)
is indebted to Prof. A. Giazotto for helpful discussions on this point).
These (quite pessimistic) conclusions hold only for the coalescence of N.S.
Let us recall that a coalescence of massive black holes 
($100 \, M_\odot$) can be detected up to 
$\sim 3{\ \rm Gpc}$.

In this paper we have showed that by phasing the signals of different 
detectors
and by using all the a priori informations given by the $\gamma$-ray burst
model,
a positive detection can be achieved in a few years of observation
time with a little improvement (less than a factor of 2) of the first 
generation of the VIRGO class detectors. 
The price to pay is the loss of information:
at the end of the observation only a positive detection of gravitational 
radiation will be achieved and a model for $\gamma$-ray burst
confirmed, but the time and the position of the events will be lost.

\end{document}